%% file: main.tex
\documentclass[a4paper, amsfonts,amssymb, amsmath, reprint, showkeys, nofootinbib, twoside, superscriptaddress]{revtex4-1}
\usepackage[english]{babel}
\usepackage[utf8]{inputenc}
\usepackage[nodisplayskipstretch]{setspace}
\usepackage[colorinlistoftodos, color=green!40, prependcaption]{todonotes}
\usepackage{amsthm}
\usepackage{mathtools}
\usepackage{xcolor}
\usepackage{graphicx}

\usepackage[margin=0.75in]{geometry}
\usepackage{adjustbox}
\usepackage{placeins}
\usepackage[T1]{fontenc}
\usepackage{lipsum}
\usepackage{csquotes}
\usepackage{algorithm2e}
\usepackage{listings}

\input{pretex}

\usepackage[pdftex, pdftitle={Article}, pdfauthor={Author}]{hyperref} 
\begin{document}
\title{Implementation of a quantum sequence alignment algorithm for quantum bioinformatics}

\author{Floyd M. Creevey}
\email{floyd.creevey@unimelb.edu.au}
\affiliation{School of Physics, University of Melbourne, VIC, Parkville, 3010, Australia.}
\author{Mingrui Jing}
\affiliation{School of Physics, University of Melbourne, VIC, Parkville, 3010, Australia.}
\author{Lloyd C. L. Hollenberg}
\email{lloydch@unimelb.edu.au}
\affiliation{School of Physics, University of Melbourne, VIC, Parkville, 3010, Australia.}

\date{\today} 

\begin{abstract}
    This paper presents the implementation of a quantum sequence alignment (QSA) algorithm on biological data in environments simulating noisy intermediate-scale quantum (NISQ) computers. The approach to quantum bioinformatics adapts the original QSA algorithm proposed in 2000 to current capabilities and limitations of NISQ-era quantum computers and uses a genetic algorithm for state preparation (GASP) to create encoding circuits to load both database and target sequences into the quantum data registers. The implementation is tested in a simulated quantum computer environment to validate the approach and refine the GASP data-loading circuit designs. The results demonstrate the practicalities of deploying the QSA algorithm and exemplify the potential of GASP for data encoding in the realm of quantum circuit design, particularly for complex algorithms in quantum bioinformatics and other data-rich problems. 
\end{abstract}
\keywords{quantum computing, genetic algorithm, quantum bioinformatics, quantum sequence alignment}

\maketitle

\section{Introduction} \label{sec:qsa_introduction}
Multiple sequence alignment (MSA) is a cornerstone of modern bioinformatics, playing a vital role in comparative genomics, evolutionary biology, and functional annotation of genes and proteins~\cite{edgar_multiple_2006}. By aligning biological sequences, be they DNA, RNA, or proteins, MSA enables researchers to identify conserved regions, infer phylogenetic relationships, and predict structural or functional motifs~\cite{chatzou_multiple_2016}. The accuracy and efficiency of sequence alignment algorithms are thus critical to the integrity of downstream analyses and the broader understanding of biological systems.

Classical algorithms such as Needleman-Wunsch for global alignment~\cite{needleman_general_1970} and Smith-Waterman for local alignment~\cite{smith_identification_1981} have long served as the gold standard in sequence alignment. These approaches utilize dynamic programming to guarantee optimal alignments. However, their computational complexity of $O(nm)$, where $n$ and $m$ are the lengths of the sequences, becomes prohibitive for large-scale biological datasets~\cite{mount_bioinformatics_2004}. Heuristic and stochastic methods have been developed to alleviate this bottleneck~\cite{durbin_biological_1998,chowdhury_review_2017}, but they often introduce trade-offs between speed and accuracy. Additionally, the alignment quality is sensitive to the choice of scoring functions and gap penalties, which are not universally optimal across biological contexts~\cite{durbin_biological_1998}.

In recent years, quantum computing has emerged as a promising alternative for accelerating computational biology tasks. Leveraging principles such as superposition and entanglement, quantum algorithms can outperform their classical counterparts in search and optimisation problems. After the advent of Grover's search algorithm~\cite{grover_fast_1996} and subsequent variants~\cite{boyer_tight_1998, brassard_quantum_1998} it was natural to consider the development of quantum approaches to sequence alignment problems. A quantum sequence alignment (QSA) algorithm utilising Grover based search to find subsequences with minimal Hamming distance was proposed in 2000~\cite{hollenberg_fast_2000}, in what appears to be one of the earliest papers in the now burgeoning area of \textit{quantum bioinformatics and biocomputing} (see~\cite{cordier_biology_2022} for a comprehensive review).  Subsequent work extended and adapted the original QSA algorithm~\cite{sarkar_qibam_2021}, and developed new quantum paradigms, including quantum annealing~\cite{nalkecz_algorithm_2022}, Fourier transform-based matching~\cite{khan_linear_2023}, and quantum-inspired evolutionary algorithms~\cite{layeb_multiple_2006,meshoul_aquantum_2005}. Interest in applying quantum computing to sequence alignment in bioinformatics is now accelerating ~\cite{kosoglu-kind_biological_2023,mongia_large_2023,sun_multiple_2012,giannakis_quantum_2019,prousalis_improving_2018,prousalis_alpha_2019}. However, aside from the as yet relatively nascent quantum hardware, the practical deployment of quantum computing for bioinformatics problems is generally hindered by the challenge of quantum state preparation, specifically, encoding classical biological data into quantum states in an efficient and hardware compatible manner.

In this paper, we present an implementation of the QSA algorithm, validating the working of the algorithm~\cite{jing_new_2021} using a Genetic Algorithm for State Preparation (GASP)~\cite{creevey_gasp_2023}, designed to generate low-depth quantum circuits that encode sequence data with tunable fidelity. Our implementation is tested in simulation, bridging the gap between theoretical quantum algorithms and their practical realisations. We demonstrate that GASP enables the preparation of quantum states with controlled fidelities, allowing QSA to operate effectively even under sub-optimal conditions. Through numerical experiments, we show that the QSA algorithm retains high alignment accuracy across a range of dataset sizes and preparation fidelities, validating its robustness and scalability.

Our results demonstrate the viability of running quantum sequence alignment algorithms on near-term quantum platforms and highlight the importance of approximate, yet efficient, state preparation techniques. This work marks a significant step toward the real-world application of quantum algorithms in bioinformatics, paving the way for future research in quantum-enhanced biological data analysis. 

The remainder of this paper will have the following structure; Section \ref{sec:QSAA} will give a summary of classical and quantum sequence alignment. Section \ref{sec:qsaa_methods} will describe the use of the GASP approach in the implementation of QSA. Section \ref{sec:qsaa_results} will present the results, and Section \ref{sec:qsaa_conclusion} the conclusions and potential future work.

\section{Classical Sequence Alignment}\label{sec:QSAA}

Sequence alignment is fundamental in bioinformatics for understanding genetic, functional, and evolutionary relationships among bio-molecular entities. Deterministic classical algorithms, such as Needleman-Wunsch~\cite{needleman_general_1970} for global alignment and Smith-Waterman~\cite{smith_identification_1981} for local alignment, are computationally intensive, especially for large datasets~\cite{mount_bioinformatics_2004}. The advent of quantum computing offers a new approach to computational biology, providing a framework for potentially alleviating these limitations. Notably, Grover's algorithm~\cite{grover_fast_1996} demonstrates a quadratic speedup in database search problems, suggesting similar benefits could be realised in sequence alignment tasks~\cite{hollenberg_fast_2000}. 

\textbf{\textit{The Needleman-Wunsch algorithm}}, developed in 1970, is a prominent algorithm for global sequence alignment. It is designed to align two sequences in their entirety, allowing for the introduction of gaps, and optimising for the best match across their lengths. The algorithm employs dynamic programming to construct a scoring matrix, systematically comparing every character of one sequence against all characters of the other, incorporating penalties for mismatches and gaps. The algorithm utilises a dynamic programming matrix, $F$, to systematically determine the optimal alignment between a query sequence,
\begin{equation}
    A = a_1 a_2 \ldots a_n,
\end{equation} 
where $n$ is the length of the query sequence, and some target sequence, 
\begin{equation}
    B = b_1 b_2 \ldots b_m,
\end{equation} 
where $m$ is the length of the target sequence. The entries of $F$, where $F(i, j)$ represents the highest score of aligning the first $i$ characters of $A$ with the first $j$ characters of $B$, are computed using a recurrence relation. 

\textbf{\textit{The Smith-Waterman algorithm}}, introduced in 1981, focuses on identifying the optimal local alignment between sequences, which is particularly useful for finding similar regions within large or diverse sequences. Like Needleman-Wunsch, this algorithm uses dynamic programming but introduces an additional condition to reset scores to zero, preventing negative scoring alignments and thereby isolating the highest-scoring local alignments. 
Both classical algorithms share a computational complexity of $O(nm)$, where $n$ and $m$ are the lengths of the two sequences being aligned. This quadratic complexity poses significant challenges as the size of the sequences or the database increases, leading to prohibitive computational demands for large-scale genomic analyses. Furthermore, the accuracy and relevance of alignments produced by these algorithms heavily depend on the choice of scoring functions and gap penalties, which are not universally optimal and require careful tuning based on the biological context~\cite{durbin_biological_1998}. Classical stochastic sequence alignment approaches have also been developed to combat the scaling of the deterministic approaches~\cite{chowdhury_review_2017}. Given sequence alignment algorithms have been so instrumental in advancing our understanding of biological sequences, their computational limitations highlight the question of the utility of quantum computers for handling the vast amounts of data generated by modern sequencing technologies. 

Quantum Sequence Alignment (QSA) leverages quantum mechanics to address the computational challenges inherent in traditional sequence alignment methodologies. Grover's algorithm~\cite{grover_fast_1996}, which provides a quadratic speedup for unstructured search problems, lays the groundwork for QSA. By encoding the search space of potential alignments into a quantum superposition, Grover's algorithm can be adapted to sift through all possible alignments simultaneously, significantly reducing the search complexity from classical's $O(nm)$ to quantum's $O(\sqrt{nm})$ for sequences of length $n$ and $m$~\cite{hollenberg_fast_2000}. Quantum walks offer another avenue for quantum-enhanced sequence alignment~\cite{kempe_quantum_2003}. By encoding sequence data onto quantum states and navigating the alignment space through quantum walks, it is possible to exploit the quantum interference patterns to efficiently converge on optimal or near-optimal alignments, potentially surpassing deterministic classical dynamic programming techniques in both speed and scalability~\cite{kempe_quantum_2003}. However, designing approaches that effectively encode biological sequences into quantum states is non-trivial. The algorithmic complexity of handling gaps, mismatches, and the biological context of alignments in a quantum framework requires innovative solutions to fully leverage quantum computing's potential. As quantum hardware continues to evolve and mature, the implementation of QSA becomes increasingly feasible. Future research will likely focus on refining quantum algorithms for sequence alignment, exploring hybrid quantum-classical approaches, and integrating QSA into broader bioinformatics workflows. The ultimate goal is to realize QSA's potential to significantly accelerate sequence alignment, unlocking new possibilities in genomics research, evolutionary studies, and personalized medicine.

\section{Quantum Sequence Alignment (QSA) Algorithm}\label{sec:qsaa_methods}

    \begin{figure*}
        \centering
        \includegraphics[width=0.99\textwidth]{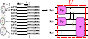}
        \caption[DNA Sequences Alignment and Encoding]{DNA Sequences Alignment and Encoding. DNA is structured as a series of bases from a basic $\{A, T, G, C\}$ alphabet, i.e $z=2$. The totality of the information contained within the DNA is the genome. Each nucleobase is represented by a bit string $\prod_{\alpha=0}^{2}B_{i, \alpha}$, encoded on 2 qubits. All consecutive sub-sequences of length $m$ ($m=3$ in this example) from a dataset $\cD$ containing $N=12$ nucleobases are encoded to produce $|\phi_0\rangle, |\phi_1\rangle, \ldots, |\phi_{N-3}\rangle$ where $\ket{\phi_i}=\bigotimes_{\alpha=0}^{2\cdot3-1}|q_{i, \alpha}\rangle$. The entire database is then represented by $|\psi_{\cD}\rangle = \frac{1}{\sqrt{N-3+1}}\sum_{i=0}^{N-3}\ket{\phi_i}\otimes\ket{i}$, encoded in a circuit by $U_{\cD}$. The target sequence $x$ containing $m$ nucleobases is encoded to produce $|\psi_x\rangle$, using $U_x$. The Hamming register is initialised with $\lceil\log_2(N + 1)\rceil$ qubits. The database and sequence registers are then linearly entangled with $CNOT$ gates, and the Hamming distances, $T[i]$, calculated by applying the $\hat{T}:\hat{T}|\bar{\phi}_i\rangle=T[i]|\bar{\phi}_i\rangle$ operator, where $T[i]=\sum_{\alpha=0}^{2\cdot 3-1}\bar{q}_{i, \alpha}$. This entire process is the initialisation unitary, $U$.}
        \label{fig:test_bio_fig}
    \end{figure*}


    Although the original framing for the QSA algorithm was in the context of protein sequences~\cite{hollenberg_fast_2000}, to facilitate the extension to DNA sequences can be recast in terms of a general database alignment problem over arbitrary alphabets, which we do here. 
    Consider a database sequence $\mathcal{D} = \{S_0, S_1,...,S_N-1\}$ of length $N$ comprising characters of the specific bio-alphabet in question (e.g. for DNA the alphabet is $S_i \in \{A, T, G, C\}$). A target sequence $x = \{s_0, s_1,...,s_{m-1}\}$ comprising $m$ characters in the same bio-alphabet is to be aligned with the database sequence. To translate to the QSA framework, the sequence characters from the $L$-letter bio-alphabet are represented as $z$-bit binary strings ($z = \lceil\log_2 L\rceil$) as: $S_i \equiv B_{i0} B_{i1}... B_{iz-1}$ and $s_i \equiv b_{i0} b_{i1}... b_{iz-1}$.
    On a quantum computer this system can be set up on two registers, the first comprised of $Q$ qubits, and the second comprised of $q$ qubits. These registers hold the encoded bit-wise representation of the sequences as,
    \begin{equation}
        \ket{\Psi_D}\equiv\frac{1}{\sqrt{N-m+1}}\sum_{i=0}^{N-m}\ket{\phi_i}\otimes\ket{i},
    \end{equation}
    where $\ket{\phi_i}$ represents the $i$th $m$ length subsequence in the database, encoded in the first register with $Q=z\cdot m$ i.e., 
    \begin{equation}
    \ket{\phi_i}\equiv\bigotimes_{\alpha=i}^{i+m-1}\bigotimes_{\beta=0}^{z-1}\ket{B_{\alpha\beta}}\equiv\bigotimes_{\alpha=0}^{z\cdot m-1} \ket{q_{i\alpha}}.
    \end{equation}
    As such, $N-m+1$ subsequences will be formed from an $N$ length database. This encoding also holds the position information of the sub-sequences in the binary numbers, $\ket{i}$, in the second register. A sequence position operator, $\hat{X}$, which acts on the Hilbert space of the index register is used to access this position information, given as,
    \begin{equation}
        \hat{X}\ket{i}=i\ket{i} \ (0\le i\le N-m),
    \end{equation}
    which forces $q$ to satisfy $2^{q}>N-m$. The final step in the initialisation process is encoding a table, $T[0, 1, \ldots, N-m]$, that holds measurements of the Hamming distances between the database states, $\ket{\phi_i}$, and the target sequence states, $x$. This is created by applying CNOT gates between each qubit in the database register and the sequence register, i.e.,
    \begin{equation}
        |\Psi_H\rangle=U_{\rm{CNOT}}(x)\ket{\Psi_D}\equiv\frac{1}{\sqrt{N-m+1}}\sum_{i=0}^{N-m}|\bar{\phi}_i\rangle\otimes\ket{i},
    \end{equation}
    where $\bar{\phi}_i$ means the opposite of $\phi_i$, and $|\bar{\phi}_i\rangle$ are the `Hamming states', 
    
    \begin{equation}
        |\bar{\phi}_i\rangle=\bigotimes_{\alpha=0}^{z\cdot m-1}|\bar{q}_{i\alpha}\rangle.
    \end{equation}
    
    An operator, $\hat{T}$ is introduced which acts on the state $|\bar{\phi}_i\rangle$ and returns $T[i]$, as,
    \begin{equation}
        \hat{T}:\hat{T}|\bar{\phi}_i\rangle=T[i]|\bar{\phi}_i\rangle,
    \end{equation}
    where,
    \begin{equation}
        T[i]=\sum_{\alpha=0}^{z\cdot m-1}\bar{q}_{i\alpha}.
    \end{equation}
    A visual representation of the initialisation process is shown in Figure \ref{fig:test_bio_fig}. 

    With initialisation completed, a search problem can be defined to find alignments of increasing Hamming distance. This is because the number of solutions, or if there is a solution, is not known at initialisation, and as such, Grover's algorithm~\cite{grover_fast_1996} cannot be used directly. However, an extension to Grover's algorithm, the BHHT algorithm~\cite{brassard_quantum_1998}, which performs a search with an unknown number of solutions, $N_t$, can be used to find a match in $O(\sqrt{\frac{N}{N_t}})$ steps. In the context of this algorithm, this requires tailoring the search to be based on all target states that satisfy $T[i]=n$, where $n$ is a Hamming distance pre-defined by the algorithm. To illustrate this, the geometric picture is very useful. Initially, the algorithm decomposes the state $|\Psi_H\rangle$ into orthogonal components with respect to the encoded target sequence, $|\bar{x}\rangle$,
    \begin{equation}
        |\Psi_H\rangle=\sqrt{\frac{N-m}{N-m+1}}|R\rangle+\frac{1}{\sqrt{N-m+1}}|S\rangle
    \end{equation}
    where, 
    \begin{equation}
        \begin{aligned}
            |S\rangle &= \frac{1}{\sqrt{N-m+1}}\sum_{i\in T[i]=n}|\bar{x}\rangle\otimes\ket{i}, \\
            |R\rangle &= \frac{1}{\sqrt{N-m}}\sum_{i\in T[i]\ne n}|\bar{\phi}_i\rangle\otimes\ket{i}.
        \end{aligned}
    \end{equation}
    The Grover operator, $G_\delta$, is constructed from the reflection operators in the Hilbert space of the first register,
    \begin{equation}
        \begin{aligned}
            I_S(n) &= 1 - 2|S\rangle\langle S|, \\
            I_H &= 1-2|\Psi_H\rangle\langle\Psi_H|.
        \end{aligned}
    \end{equation}
    The action of $I_S$, in the original framework of Grover's algorithm, is to implement the oracle function, $F(i)$, as a phase shift to the state $|\bar{\phi}_i\rangle$ depending on the search criteria $T[i]=n$,
    \begin{equation}
        I_S(n)|\bar{\phi}_i\rangle=(-1)^{F(i)} \ket{\bar{\phi}_i}=\begin{cases}
                    -|\bar{\phi}_i\rangle \ \text{if $T[i]=n$} \\
                    |\bar{\phi}_i\rangle \ \text{otherwise}.
                \end{cases}
    \end{equation}
    The BHHT algorithm is applied at each iteration, with a repeat index $r$ as a pre-determined measure of the search confidence level. Overall, the QSA algorithm can be described as follows,
    \begin{enumerate}
        \item $0$th iteration: search for an occurrence of a state with $T[i]=0$. If successful measure the position and exit, if unsuccessful after $r$ repeats go to the next iteration
        \item $n$th iteration: search for an occurrence of a state with $T[i]=n$ using $U = -I_HI_S(n)$. If successful measure the position and exit, if unsuccessful after $r$ repeats go to the next iteration, setting $n\to n+1$
        \item Exit at some iteration $n=k$ where one optimal alignment $T[i_k]=k$ and its position $i_k$ has been found
    \end{enumerate}
    Current technology quantum computers are not yet advanced enough for systems the size of realistic biological data, so this work will focus on proof of concept on 5-10 qubit systems. Building on the work done in~\cite{jing_new_2021}, some simplifications were made to the original QSA algorithm to aid in implementation. The position register in the original algorithm could be removed by reversing the database and sample registers in the circuit, changing the database wave function to,
    \begin{equation}
        |\psi_D\rangle = \frac{1}{\sqrt{N}}\sum_{i=1}^N|D_i\rangle,
    \end{equation}
    where $D_i$ are the binary strings of the characters of a database containing $N$ sequences. As the maximum Hadamard distance from any $n$ length string is $n + 1$, and an $n$ length register is required to encode both the database sequences and target sequence, as such the entire algorithm requires $2n + k$ qubits, where $k=\lceil\log_2(n + 1)\rceil$. The total state after encoding the database sequences, $|\psi_D\rangle$ and target sequence, $|S\rangle$, is then denoted,
    \begin{equation}
        |\psi_{\rm{tot}}\rangle = |\psi_D\rangle|S\rangle|0\rangle^{\otimes k}.
    \end{equation}
    The database and sample registers are then linearly entangled with CNOT gates altering the total state to,
    \begin{equation}
        \begin{aligned}
            U_{\rm{CNOT}}|\psi_{\rm{tot}}\rangle &= \frac{1}{\sqrt{N}}\sum_{i=1}^N|D_i\rangle|S\oplus D_i\rangle|0\rangle^{\oplus k} \\
            &= \frac{1}{\sqrt{N}}\sum_{i=1}^N|D_i\rangle|\bar{S_i}\rangle|0\rangle^{\oplus k}.
        \end{aligned}
    \end{equation}
    The $\hat{T}$ operator to determine the Hamming distances, $d_i$, and store them in a quantum state in the Hamming register then becomes,
    \begin{equation}
        \hat{T}|\bar{S_i}\rangle|0\rangle^{\otimes k} = |\bar{S_i}\rangle|d_i\rangle.
    \end{equation}

    \begin{figure*}
        \centering
        \includegraphics[width=0.99\textwidth]{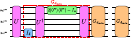}
        \caption[Quantum Search Algorithm Circuit]{Quantum search algorithm circuit. The circuit is comprised much like Grover's algorithm, with repeating Grover layers. In this case, the Grover layer, $G_{d_{\rm{min}}}$, is composed of the query to the oracle, $I_\delta$, the inverse database unitary, $U^\dag$, inversion about the mean, and finally the database unitary, $U$.}
        \label{fig:qsaa_flow}
    \end{figure*}
    
    In practice, $\hat{T}$ comprises a series of multi-controlled $X$ gates. All possible $2^i$ controlled $X$ gates are applied for each $i$th qubit in the Hamming register to the qubits in the sample register (where $i$ is indexed from the bottommost qubit of $k$ to the topmost). With all these operations completed, the total state is,
    \begin{equation}
        |\psi_{\rm{tot}}'\rangle = \frac{1}{\sqrt{N}}\sum_{i=1}^N|D_i\rangle|\bar{S_i}\rangle|d_i\rangle,
    \end{equation}
    encoded by the initialisation unitary, $U$. As a perfect match will not always exist in the database, we require a match with the minimum Hamming distance, $d_{\rm{min}}$, of which there could be multiple. Assuming there are multiple of these minimum matches, $M$, in the database, where the number of matches is less than the size of the database, $M < N$, $|\psi_{\rm{tot}}'\rangle$ can be decomposed into orthogonal components form,
    \begin{equation}
        |\psi_{\rm{tot}}'\rangle = \sqrt{\frac{N-M}{N}}|\alpha\rangle + \sqrt{\frac{M}{N}}|\beta\rangle,
    \end{equation}
    where,
    \begin{eqnarray}
        |\alpha\rangle = \frac{1}{\sqrt{N-M}}\sum_{i\ne i_\beta}|D_i\rangle\otimes|\bar{S}_i\rangle\otimes|d_i\rangle, \\
        |\beta\rangle = \frac{1}{\sqrt{M}}\sum_{i_\beta\in D_\beta}|D_{i_{\beta}}\rangle\otimes|\bar{S}_{i_{\beta}}\rangle\otimes|d_{\min}\rangle. \\
    \end{eqnarray}
    The operative register in this algorithm is the database register, where the best match is found. The phase-inversion and mean-inversion gates are expressed as \textit{reflection} operators,
    \begin{eqnarray}
        I_{d_{\min}} = I - 2|d_{\min}\rangle\langle d_{\min}|, \\
        I_D = 2|\psi_{\rm{tot}}'\rangle\langle\psi_{\rm{tot}}'| - I.
    \end{eqnarray}
    $I_{d_{\min}}$ holds the query to the oracle function $F(i)$ and acts on the Hamming register to give the phase shift,
    \begin{equation}
        I_{d_{\min}}|d_i\rangle = (-1)^{F(i)}|d_i\rangle = \begin{cases}
            -|d_i\rangle & \text{if}\ d_i=d_{\min} \\
            |d_i\rangle & \ \text{otherwise}.
        \end{cases}
    \end{equation}
    These reflection operators can then be used to construct the Grover operator $G_{d_{\min}}$ for the distance $d_{\min}$, for each step of the algorithm, 
    \begin{equation}
        G_{d_{\min}} = I_DI_{d_{\min}}
    \end{equation}
    The function of the algorithm is to calculate $d_{\min}$ by iterating through increasing Hamming distances. This is achieved by modifying the oracle by changing $I_{d_{\min}}$ to $I_{\delta}$ where,
    \begin{equation}
        I_{\delta}|d_i\rangle = \begin{cases}
            -|d_i\rangle & \text{if}\ d_i=\delta \\
            |d_i\rangle & \text{if}\ \text{otherwise}.
        \end{cases}
    \end{equation}
    
    A visual representation of the circuit structure for this algorithm is shown in Figure \ref{fig:qsaa_flow}, and pseudocode of the complete algorithmic procedure is presented in Algorithm \ref{alg:GASP_QSAA}.

    \RestyleAlgo{ruled}
    \SetKwComment{Comment}{/* }{ */}

    \begin{algorithm*}
    \scriptsize
        \caption{QSA with GASP Initialisation}\label{alg:GASP_QSAA}
        \KwData{$data$, $targetSequence$}
        \KwResult{$minimumDistance \ (int)$, $minimumDistanceSequence \ (string)$}
        $hammingDistance \gets 0$\;
        $dataQubits, \ dataRegister \gets GASP(data)$
        
        \While{$hammingDistance \le dataQubits$}{
            $matches \gets calculateMatches(hammingDistance)$\;
            $\theta_g \gets \sin(\frac{\sqrt{length(matches)}}{\sqrt{span(data)}})$\;
            $iterations \gets \frac{\pi}{4}\sqrt{\frac{span(data)}{length(matches)}}$\;
            $circuit \gets QuantumCircuit()$\;
            \For{$i \gets 0$ \KwTo $iterations$}{
                $circuit += goverIteration()$\;
            }
            $result = execute(circuit)$\;
            \eIf(){$result\in matches$}{
                \textbf{Return} $result$\;
            }{
                $hammingDistance += 1$
            }
        }
    \end{algorithm*}

    As there can be multiple solutions to QSA, for example, if there is no $0$ Hamming distance match, but multiple $1$ Hamming distance matches, it is important to describe how accuracy is calculated. 
    Let $\hat{u}$ and $\hat{v}$ be two probability distributions over $b = 2^k$ computational basis states of a $k$-qubit quantum system, where $\hat{u}$ is obtained from a counts dictionary (e.g., a histogram of quantum measurement outcomes), normalised such that $\sum_{i=1}^b u_i = 1$ and $\hat{v}$ is the ideal probability distribution obtained from the squared amplitudes of the theoretical (ideal) quantum statevector.    
    Each component of the vectors is defined as,
    \begin{equation}
        u_i = \frac{\text{counts}[i]}{\sum_{j=1}^b \text{counts}[j]}, \quad 
        v_i = |\langle i \! \mid \! \psi \rangle|^2.
    \end{equation}
    We define accuracy using the cosine similarity (CS) between $\hat{u}$ and $\hat{v}$ as given by,
    \begin{equation}
    \begin{aligned}
        \text{CS}(\hat{u}, \hat{v}) =
        \frac{\hat{u} \cdot \hat{v}}{\|\hat{u}\| \cdot \|\hat{v}\|} 
    \end{aligned}
    \end{equation}  
    where the $\|\cdot\|$ denotes the vector $2$-norm. This metric measures the cosine of the angle between the two vectors in the $n$-dimensional space. It ranges from $0$ (orthogonal) to $1$ (identical direction).

\section{Implementation of QSA with GASP Data Encoding}\label{sec:qsaa_results}

    \begin{figure*}
        \centering
        \includegraphics[width=\linewidth]{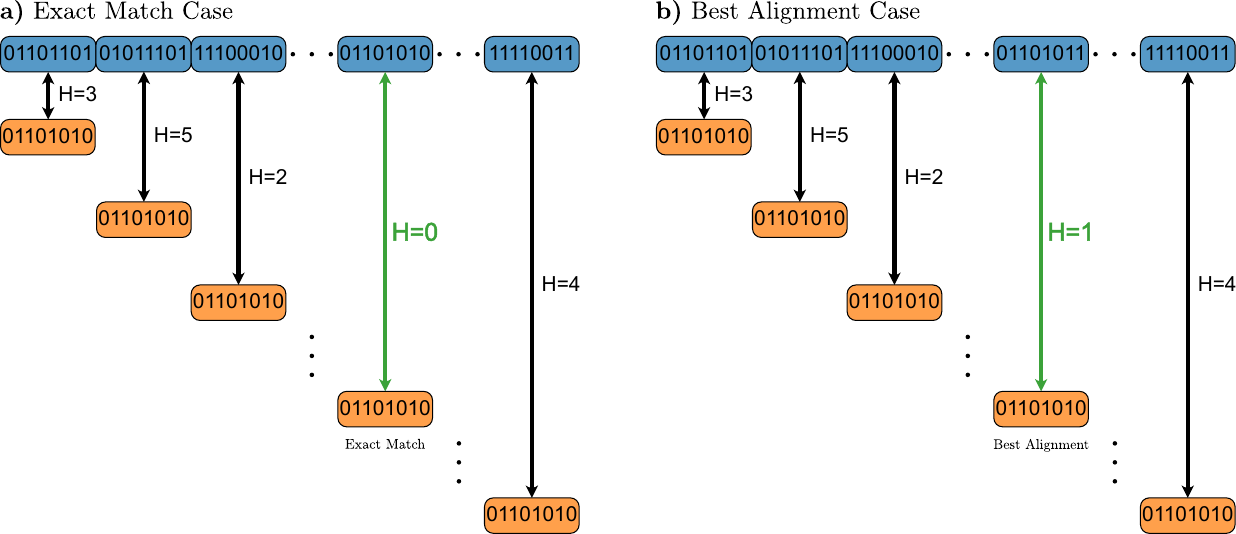}
        \caption{Comparison between perfect match and best alignment cases. The QSA algorithm searches through a database for an entry of a given Hamming distance from the given target. If the given target exists in the database, i.e the Hamming distance, $H$, is $0$, then an exact match is found. Otherwise, the algorithm will continue to search through the database for entries with larger Hamming distances from the target, finding the best alignment, or smallest Hamming distance match.}
        \label{fig:qsaa_alignment_figure}
    \end{figure*}

\begin{figure*}
        \centering
        \includegraphics[width=0.99\textwidth]{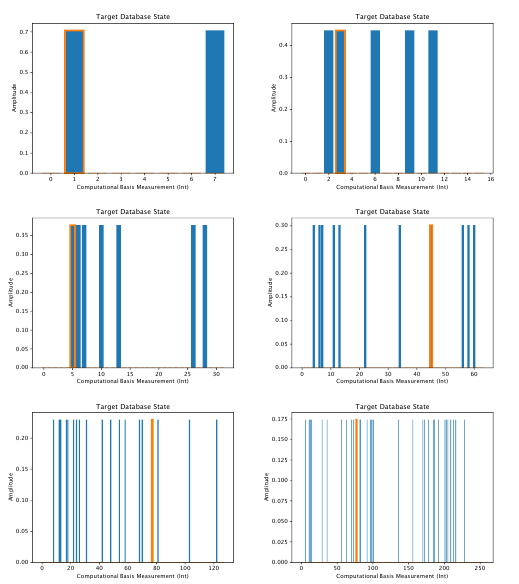}
        \caption[Example Databases for QSA]{Example uniform amplitude databases for QSA. Randomly generated databases of $3$, $4$, $5$, $6$, $7$, and $8$ qubits are shown. Each database contains a random set of $N = \lceil \frac{2^n}{n} \rceil$ sequences in an even superposition, coloured blue. The randomly chosen target sequence is coloured orange. Note: in this example the target was always in the database.}
        \label{fig:random_databases}
    \end{figure*}

    In implementation, we simplify to consider alignment with respect to small-scale binary string examples, as displayed in Figure \ref{fig:qsaa_alignment_figure}, commensurate with current NISQ level hardware. Additionally, to compress the initialisation depth we use the genetic algorithm for state preparation (GASP)~\cite{creevey_gasp_2023} approach to encode the database sequences $B_0 B_1...B_N$ for $N = 3\ldots8\ldots$ and then continue. QSA was run using GASP to prepare the data register states of $3-8$ qubits over a range of preparation fidelities, $1 - \epsilon$. These finite fidelity data register states, $|\psi_{G}(\epsilon)\rangle$, were created by altering the ideal data register state, $|\psi_t\rangle$, in the following manner,
    \begin{equation}
        |\psi_{G}(\epsilon)\rangle = e^{-i\epsilon \hat{H}}|\psi_t\rangle,
    \end{equation}
    where $\hat{H}$ is some random $2^n\times2^n$ ($n$ is the number of qubits in the data register) Hermitian matrix, and $\epsilon$ is a parameter such that $|\langle\psi_t|\psi_{G}(\epsilon)\rangle|^2$ is equal to the a priori fidelity ($\epsilon$ was in practice determined via a classical optimiser). Creating these lower fidelity data register states ensures the accurate representation of the fidelities in the states generated by GASP. As genetic algorithms are stochastic, when given a target fidelity a state with a much higher fidelity than that of the target fidelity could be generated. By determining these lower fidelity data registers, GASP could be tasked to generate them to $99\%$ fidelity, ensuring the registers generated were of the desired fidelities. For each number of qubits in the data register, a random database with sequences in an even superposition was generated, together with a random target sequence corresponding to the bit string representation of integers $1-2^n$. Each of these databases contained $\lceil \frac{2^n}{n} \rceil$ sequences, to maintain the proportion of elements across tests of different numbers of qubits. Some examples of these generated databases are shown in Figure \ref{fig:random_databases}.

    For each database of a given fidelity, $|\psi_{G}(\epsilon)\rangle$, the number of possible matches, $c$, starting from a Hamming distance of $0$, for the target sequence in the given database, was calculated. These two parameters allowed the calculation of the optimal number of Grover operators, $p$, in the QSA circuit, $p_{\rm{opt}}$,
    \begin{equation}
        p_{\rm{opt}} = \left\lceil \frac{\pi}{4}\sqrt{\frac{N}{c}} \right \rceil.
    \end{equation}

    \begin{figure*}
        \centering
        \includegraphics[width=\textwidth]{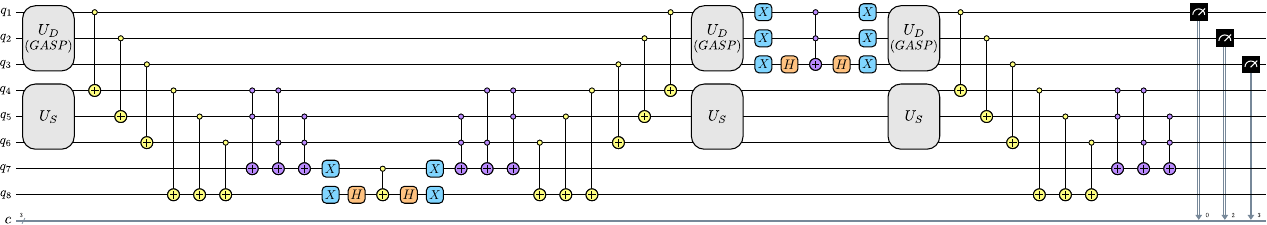}
        \caption[Example QSA Circuit]{Example QSA circuit for a data register size of 3 qubits. The QSA circuit is a quantum circuit with $2n + \lceil\log_2(n+1)\rceil$ qubits. To construct the circuit $U$ is applied, followed by $p$ layers of $G_{d_{\rm{min}}}$.}
        \label{fig:QSAA_circ}
    \end{figure*}

\begin{figure*}
        \centering
        \includegraphics[width=0.99\textwidth]{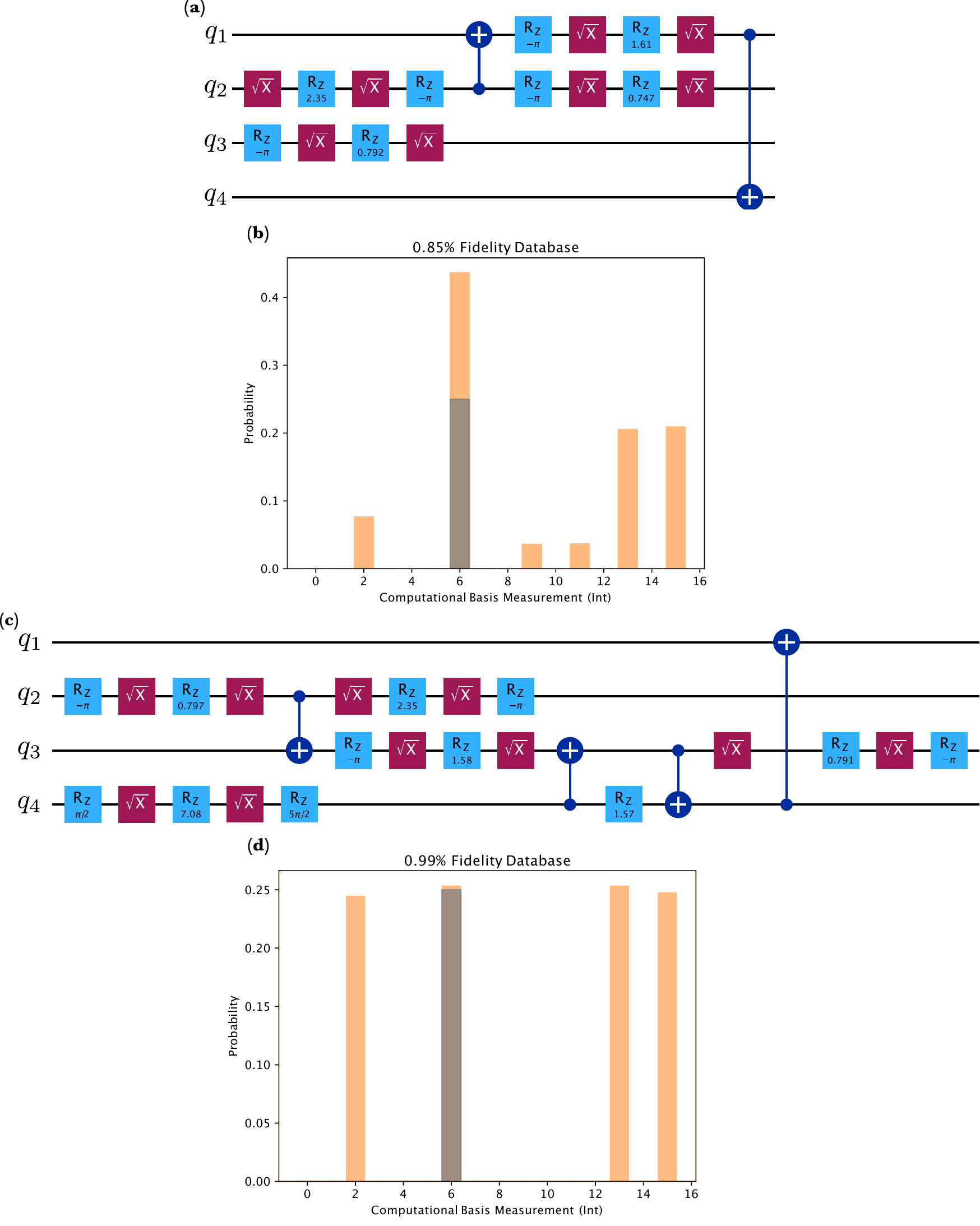}
        \caption[Example Lower Fidelity States and Corresponding GASP Generated Circuits]{Example lower fidelity states and corresponding GASP generated circuits. \textbf{(a)} shows the GASP generated circuit for a $99\%$ fidelity $4$ qubit state with $\lceil \frac{2^n}{n} \rceil$ active amplitudes. The circuit contained $26$ gates, $4$ CNOT gates, and had a depth of $17$. \textbf{(b)} shows the probability distribution for the $99\%$ fidelity $4$ qubit state. The resulting distribution is shown in blue, and the target is shown in orange. \textbf{(c)} shows the GASP generated circuit for a $85\%$ fidelity $4$ qubit state with $\lceil \frac{2^n}{n} \rceil$ active amplitudes. The circuit contained $18$ gates, $2$ CNOT gates, and had a depth of $10$. \textbf{(d)} shows the probability distribution for the $85\%$ fidelity $4$ qubit state. The resulting distribution is shown in orange, and the target is shown in grey.}
        \label{fig:lower_fidelity_states}
    \end{figure*}

    An example circuit for QSA for a 3 qubit data register with 1 layer, displaying the construction of the $\hat{T}$, $I_{\delta}$, and $2|0^n\rangle\langle0^n|-I_n$ operators, is shown in Figure \ref{fig:QSAA_circ}.
    Examples of GASP-generated circuits and probability distributions for $99\%$ and $95\%$ fidelity $4$ qubit states with $\lceil \frac{2^n}{n} \rceil$ active amplitudes can be seen in Figure \ref{fig:lower_fidelity_states}.

\begin{figure*}
        \centering
        \includegraphics[width=0.99\textwidth]{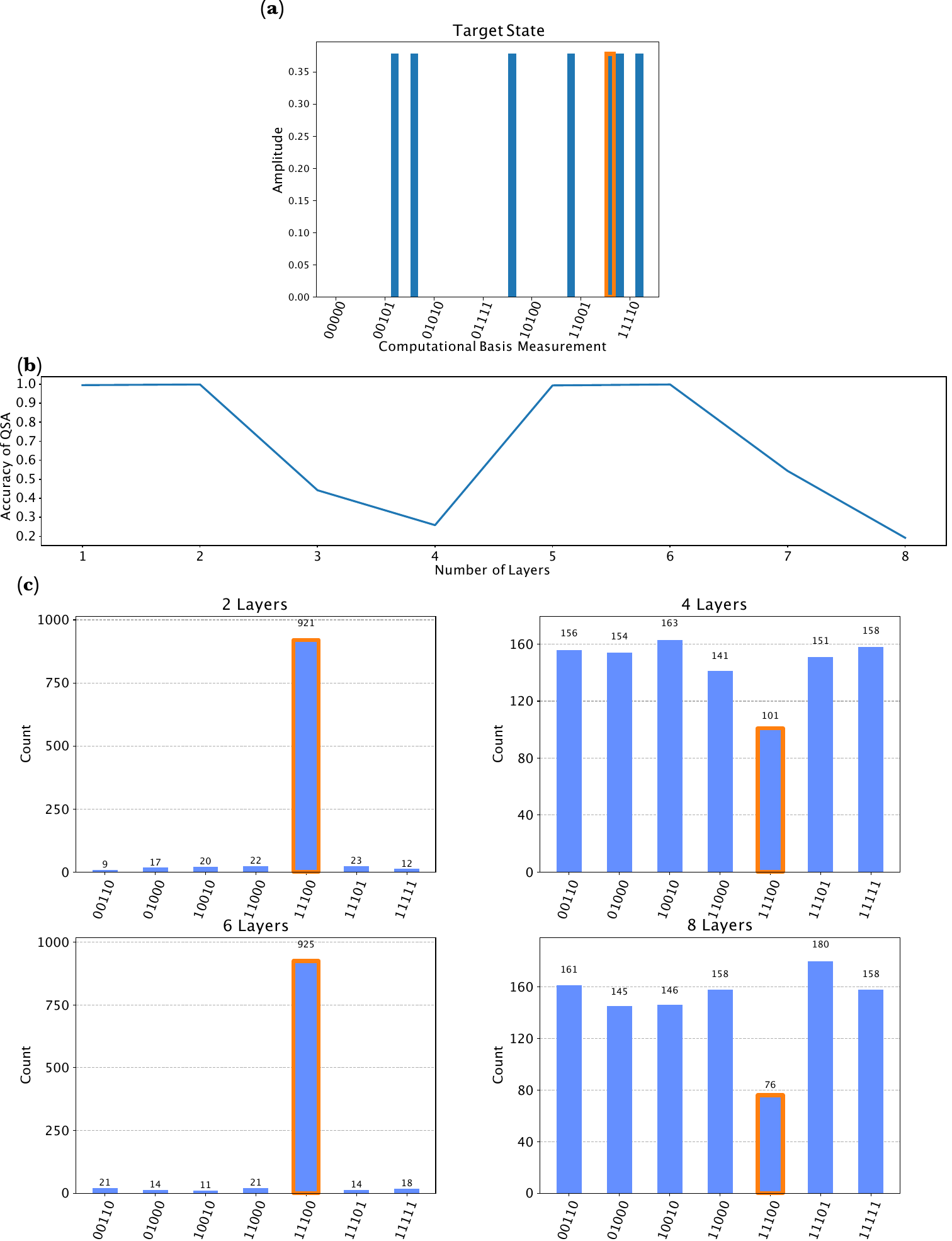}
        \caption[QSA Accuracy with Various Layers]{QSA accuracy with respect to the number of Grover layers in the QSA circuit. \textbf{(a)} shows the loaded data register with the target sequence outlined in orange. \textbf{(b)} shows the accuracy of QSA plotted against the number of layers in the QSA circuit. It can be seen that the accuracy of QSA cycles from optimal at $p=1$ and $p=2$, and sub-optimal at $p=3$ and $p=4$, indicating optimality every $4p$.\textbf{(c)} shows the resultant distributions at 2, 4, 6, and 8 layers with the target sequence outlined in orange.}
        \label{fig:QSAA_cycle}
    \end{figure*}

    The number of layers used has a large impact on the final accuracy of QSA. If a sub-optimal number of layers is used, the algorithm could cause the rotation from the initial state to the desired state to be too little or too great. Figure \ref{fig:QSAA_cycle} shows how the accuracy of QSA varies as the number of layers varies. This is a $5$ qubit example with the target string existing in the database, so $c=1$, and $N=\lceil\frac{2^5}{5}\rceil=7$. It can be seen that the accuracy of QSA cycles from optimal at 1-2 layers, and sub-optimal at 3-4 layers, indicating optimality every 4 layers as one expects from the geometric picture of Grover's algorithm. The circuit was then executed, and the counts returned. If the sequence with the maximum number of counts was not in $M$, the Hamming distance was incremented by 1, and the algorithm was re-run. 

    \begin{figure*}
        \centering
        \includegraphics[width=0.99\textwidth]{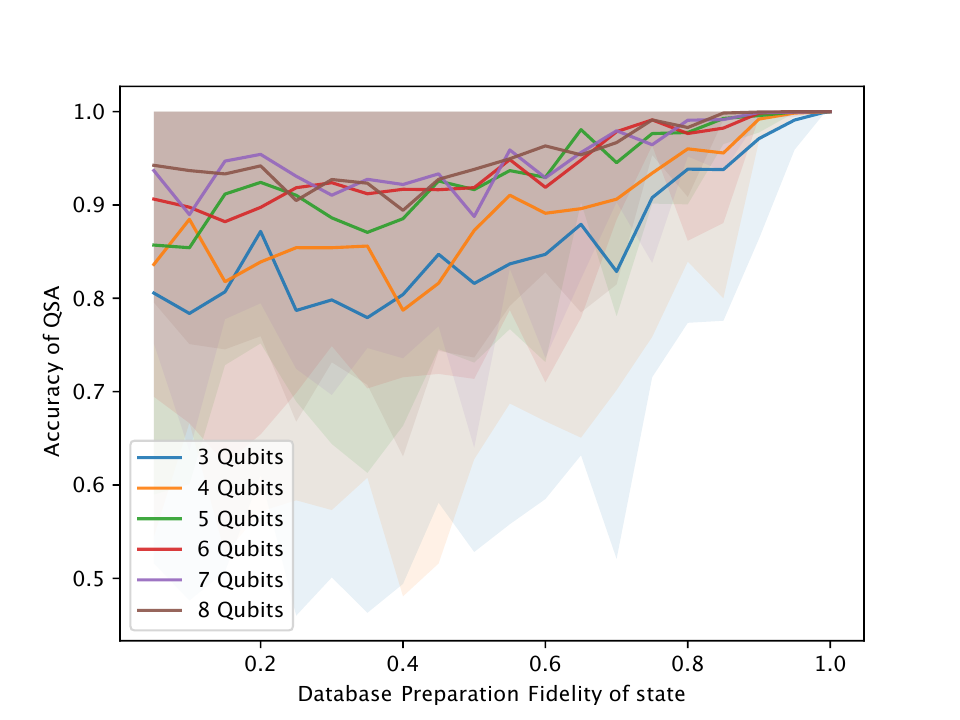}
        \caption[Accuracy of QSA Over Various Data Register Preparation Fidelities]{Accuracy of QSA over various data register preparation fidelities. The results of QSA with $3$, $4$, $5$, $6$, $7$, and $8$ qubits, databases containing $2$, $4$, $6$, $10$, $18$, and $32$ strings, are shown in blue, orange, green, red, purple, and brown respectively. For each data register size, QSA was run with data register preparation fidelities of $5\%-100\%$ in increments of $5\%$, with the average plotted as a solid line, and the standard deviation plotted as a shaded region around the average. Note: Based on randomised $2^{n/2}$ amplitudes in the database, these results show a mix of exact match and best alignment cases.}
        \label{fig:acc_vs_fid}
    \end{figure*}

    The results were averaged for each data register size, at each fidelity, over $10$ tests. It can be seen from the results of these tests, shown in Figure \ref{fig:acc_vs_fid}, that the average accuracy of QSA overall data register sizes remains fairly high, barely dropping below $80\%$, even at very low preparation fidelities, however, the standard deviation varies wildly. However, at preparation fidelities of $80\%$ the standard deviation for all data register sizes was above $80\%$. Further, the standard deviation of the results seems to indicate a linear relationship between the accuracy of QSA and the preparation fidelity of the data. This gives great evidence to support approximate preparation algorithms, such as GASP, for these tasks. As higher required fidelities for GASP require deeper circuits, this reduction in required preparation fidelity provides evidence that GASP could aid in running QSA on current NISQ-era hardware. 

\section{Conclusion} \label{sec:qsaa_conclusion}

    This study assessed the application of GASP in the preparation of data registers for quantum sequence alignment. Our investigation delves into the performance of QSA under varying conditions of data register preparation fidelity, facilitated by GASP, to understand its robustness and accuracy in sequence alignment tasks within a quantum computing framework. The results show that QSA maintains a relatively high level of accuracy, even when the fidelity of the prepared data registers, as facilitated by GASP, is not optimal, as shown in Figure \ref{fig:acc_vs_fid}. This resilience identifies the potential of QSA for application in bioinformatics applications where accurate sequence alignment is crucial. The ability of QSA to produce reliable results, despite variations in the state preparation fidelity, shows a potential advantage for quantum computing algorithms in managing and compensating for errors and imperfections in quantum state preparation. They also demonstrate the effectiveness of GASP as a method for preparing quantum states in a manner that is both efficient and adaptable to quantum hardware. This adds evidence for the potential of quantum computing in bioinformatics, potentially surpassing the capabilities of classical computing methods in handling large-scale sequence data. Additionally, the resilience of QSA to variations in state preparation fidelity, as facilitated by GASP, encourages further exploration into quantum error mitigation strategies. This is critical for the broader adoption and application of quantum algorithms.

    This work shows the viability of using GASP for the preparation of data registers in QSA and highlights the robustness of QSA against the challenges of quantum state preparation fidelity. As quantum computing advances, the integration of algorithms like QSA and state preparation methods like GASP will play an important role in harnessing the full potential of quantum computing for bioinformatics and state preparation. Future work will focus on refining these algorithms to work efficiently and effectively on larger systems, and exploring their application to a broader range of bioinformatics challenges.

  \section{Acknowledgements} \label{sec:acknowledgements}
    This research was supported by the University of Melbourne through the establishment of the IBM Quantum Network Hub and the Australian Research Council Centre of Excellence for Quantum Biotechnology (CE230100021) at the University. FMC was supported by Australian Government Research Training Program Scholarships. This research was supported by The University of Melbourne’s Research Computing Services and the Petascale Campus Initiative. M. J. contributed to the project while at the School of Physics, University of Melbourne, VIC, Parkville, 3010, Australia, and has a current address of the Thrust of Artificial Intelligence, Information Hub, The Hong Kong University of Science and Technology (Guangzhou), China.

    \section{Author contributions statement}
        L.C.L.H. conceived the project with input from F.M.C. and M.J. The computational framework was created by F.M.C. and M.J. who also performed the experimental calculations, with input from all authors. All authors had input in writing the manuscript.
    
    \section{Competing interests} \label{sec:interests}
        The authors declare no competing interests.
    
    \section{Data availability} \label{sec:data}
        The datasets generated during and/or analysed during the current study are available from the corresponding author on reasonable request.

\bibliography{main.bib}
\clearpage

\end{document}

%% file: pretex.tex
\newcommand{\nc}{\newcommand}
\nc{\rnc}{\renewcommand}
\nc{\lbar}[1]{\overline{#1}}
\nc{\bra}[1]{\langle#1|}
\nc{\ket}[1]{|#1\rangle}
\nc{\ketbra}[2]{|#1\rangle\!\langle#2|}
\nc{\braket}[2]{\langle#1|#2\rangle}

\nc{\proj}[1]{| #1\rangle\!\langle #1 |}
\nc{\avg}[1]{\langle#1\rangle}
\nc{\liebracket}[1]{\left\langle #1\right\rangle_{\Op{Lie}}}
\nc{\smfrac}[2]{\mbox{$\frac{#1}{#2}$}}
\nc{\tr}{\operatorname{Tr}}
\nc{\ox}{\otimes}
\nc{\dg}{\dagger}
\nc{\dn}{\downarrow}

\nc{\cA}{{\mathcal A}}
\nc{\cB}{{\mathcal B}}
\nc{\cC}{{\mathcal C}}
\nc{\cD}{{\mathcal D}}
\nc{\cE}{{\mathcal E}}
\nc{\cF}{{\mathcal F}}
\nc{\cG}{{\mathcal G}}
\nc{\cH}{{\mathcal H}}
\nc{\cI}{{\mathcal I}}
\nc{\cJ}{{\mathcal J}}
\nc{\cK}{{\mathcal K}}
\nc{\cL}{{\mathcal L}}
\nc{\cM}{{\mathcal M}}
\nc{\cN}{{\mathcal N}}
\nc{\cO}{{\mathcal O}}
\nc{\cP}{{\mathcal P}}
\nc{\cQ}{{\mathcal Q}}
\nc{\cR}{{\mathcal R}}
\nc{\cS}{{\mathcal S}}
\nc{\cT}{{\mathcal T}}
\nc{\cU}{{\mathcal U}}
\nc{\cV}{{\mathcal V}}
\nc{\cX}{{\mathcal X}}
\nc{\cY}{{\mathcal Y}}
\nc{\cZ}{{\mathcal Z}}
\nc{\cW}{{\mathcal W}}
\nc{\csupp}{{\operatorname{csupp}}}
\nc{\qsupp}{{\operatorname{qsupp}}}
\nc{\var}{{\operatorname{var}}}
\nc{\rar}{\rightarrow}
\nc{\lrar}{\longrightarrow}
\nc{\polylog}{{\operatorname{polylog}}}
\nc{\wt}{{\operatorname{wt}}}
\nc{\av}[1]{{\left\langle{#1} \right\rangle}}
\nc{\supp}{{\operatorname{supp}}}
\nc{\spn}{{\operatorname{span}}}

\nc{\RR}{{{\mathbb R}}}
\nc{\CC}{{{\mathbb C}}}
\nc{\FF}{{{\mathbb F}}}
\nc{\NN}{{{\mathbb N}}}
\nc{\ZZ}{{{\mathbb Z}}}
\nc{\PP}{{{\mathbb P}}}
\nc{\QQ}{{{\mathbb Q}}}
\nc{\UU}{{{\mathbb U}}}
\nc{\EE}{{{\mathbb E}}}
\nc{\id}{{\operatorname{id}}}